\documentclass[pra,aps,nopacs,onecolumn,twoside,superscriptaddress]{revtex4}

\usepackage{amsmath,amsfonts,amssymb,caption,color,epsfig,graphics,graphicx,hyperref,latexsym,mathrsfs,revsymb,theorem,url,verbatim,epstopdf,mathtools,enumerate}
\usepackage{subfigure}

\hypersetup{colorlinks,linkcolor={blue},citecolor={blue},urlcolor={red}}

\newtheorem{definition}{Definition}
\newtheorem{proposition}[definition]{Proposition}
\newtheorem{lemma}[definition]{Lemma}

\newtheorem{theorem}[definition]{Theorem}
\newtheorem{corollary}[definition]{Corollary}
\newtheorem{conjecture}[definition]{Conjecture}

\newtheorem{remark}[definition]{Remark}
\newtheorem{example}[definition]{Example}
\newtheorem{question}[definition]{Question}
\newtheorem{memo}[definition]{Memo}

\def\squareforqed{\hbox{\rlap{$\sqcap$}$\sqcup$}}
\def\qed{\ifmmode\squareforqed\else{\unskip\nobreak\hfil
\penalty50\hskip1em\null\nobreak\hfil\squareforqed
\parfillskip=0pt\finalhyphendemerits=0\endgraf}\fi}
\def\endenv{\ifmmode\;\else{\unskip\nobreak\hfil
\penalty50\hskip1em\null\nobreak\hfil\;
\parfillskip=0pt\finalhyphendemerits=0\endgraf}\fi}
\newenvironment{proof}{\noindent \textbf{{Proof.~} }}{\qed}
\def\Dbar{\leavevmode\lower.6ex\hbox to 0pt
{\hskip-.23ex\accent"16\hss}D}
\makeatletter
\def\url@leostyle{%
  \@ifundefined{selectfont}{\def\UrlFont{\sf}}{\def\UrlFont{\small\ttfamily}}}
\makeatother
\def\bcj{\begin{conjecture}}
\def\ecj{\end{conjecture}}
\def\bcr{\begin{corollary}}
\def\ecr{\end{corollary}}
\def\bd{\begin{definition}}
\def\ed{\end{definition}}
\def\bea{\begin{eqnarray}}
\def\eea{\end{eqnarray}}
\def\beq{\begin{equation}}
\def\eeq{\end{equation}}
\def\bal{\begin{aligned}}
\def\eal{\end{aligned}}
\def\bem{\begin{enumerate}}
\def\eem{\end{enumerate}}
\def\bex{\begin{example}}
\def\eex{\end{example}}
\def\bim{\begin{itemize}}
\def\eim{\end{itemize}}
\def\bl{\begin{lemma}}
\def\el{\end{lemma}}
\def\bma{\begin{bmatrix}}
\def\ema{\end{bmatrix}}
\def\bpf{\begin{proof}}
\def\epf{\end{proof}}
\def\bpp{\begin{proposition}}
\def\epp{\end{proposition}}
\def\bqu{\begin{question}}
\def\equ{\end{question}}
\def\br{\begin{remark}}
\def\er{\end{remark}}
\def\bt{\begin{theorem}}
\def\et{\end{theorem}}
\def\bmm{\begin{memo}}
\def\emm{\end{memo}}

\def\btb{\begin{tabular}}
\def\etb{\end{tabular}}

\newcommand{\nc}{\newcommand}

\def\a{\alpha}

\def\i{\iota}

\def\s{\sigma}

\def\ph{\varphi}

\def\og{\omega}

\nc{\bbA}{\mathbb{A}} \nc{\bbB}{\mathbb{B}} \nc{\bbC}{\mathbb{C}}
 \nc{\bbD}{\mathbb{D}} \nc{\bbE}{\mathbb{E}} \nc{\bbF}{\mathbb{F}}
 \nc{\bbG}{\mathbb{G}} \nc{\bbH}{\mathbb{H}} \nc{\bbI}{\mathbb{I}}
 \nc{\bbJ}{\mathbb{J}} \nc{\bbK}{\mathbb{K}} \nc{\bbL}{\mathbb{L}}
 \nc{\bbM}{\mathbb{M}} \nc{\bbN}{\mathbb{N}} \nc{\bbO}{\mathbb{O}}
 \nc{\bbP}{\mathbb{P}} \nc{\bbQ}{\mathbb{Q}} \nc{\bbR}{\mathbb{R}}
 \nc{\bbS}{\mathbb{S}} \nc{\bbT}{\mathbb{T}} \nc{\bbU}{\mathbb{U}}
 \nc{\bbV}{\mathbb{V}} \nc{\bbW}{\mathbb{W}} \nc{\bbX}{\mathbb{X}}
 \nc{\bbZ}{\mathbb{Z}}

 \nc{\bA}{{\bf A}} \nc{\bB}{{\bf B}} \nc{\bC}{{\bf C}}
 \nc{\bD}{{\bf D}} \nc{\bE}{{\bf E}} \nc{\bF}{{\bf F}}
 \nc{\bG}{{\bf G}} \nc{\bH}{{\bf H}} \nc{\bI}{{\bf I}}
 \nc{\bJ}{{\bf J}} \nc{\bK}{{\bf K}} \nc{\bL}{{\bf L}}
 \nc{\bM}{{\bf M}} \nc{\bN}{{\bf N}} \nc{\bO}{{\bf O}}
 \nc{\bP}{{\bf P}} \nc{\bQ}{{\bf Q}} \nc{\bR}{{\bf R}}
 \nc{\bS}{{\bf S}} \nc{\bT}{{\bf T}} \nc{\bU}{{\bf U}}
 \nc{\bV}{{\bf V}} \nc{\bW}{{\bf W}} \nc{\bX}{{\bf X}}
 \nc{\bZ}{{\bf Z}}

\nc{\cA}{{\cal A}} \nc{\cB}{{\cal B}} \nc{\cC}{{\cal C}}
\nc{\cD}{{\cal D}} \nc{\cE}{{\cal E}} \nc{\cF}{{\cal F}}
\nc{\cG}{{\cal G}} \nc{\cH}{{\cal H}} \nc{\cI}{{\cal I}}
\nc{\cJ}{{\cal J}} \nc{\cK}{{\cal K}} \nc{\cL}{{\cal L}}
\nc{\cM}{{\cal M}} \nc{\cN}{{\cal N}} \nc{\cO}{{\cal O}}
\nc{\cP}{{\cal P}} \nc{\cQ}{{\cal Q}} \nc{\cR}{{\cal R}}
\nc{\cS}{{\cal S}} \nc{\cT}{{\cal T}} \nc{\cU}{{\cal U}}
\nc{\cV}{{\cal V}} \nc{\cW}{{\cal W}} \nc{\cX}{{\cal X}}
\nc{\cZ}{{\cal Z}}

\nc{\hA}{{\hat{A}}} \nc{\hB}{{\hat{B}}} \nc{\hC}{{\hat{C}}}
\nc{\hD}{{\hat{D}}} \nc{\hE}{{\hat{E}}} \nc{\hF}{{\hat{F}}}
\nc{\hG}{{\hat{G}}} \nc{\hH}{{\hat{H}}} \nc{\hI}{{\hat{I}}}
\nc{\hJ}{{\hat{J}}} \nc{\hK}{{\hat{K}}} \nc{\hL}{{\hat{L}}}
\nc{\hM}{{\hat{M}}} \nc{\hN}{{\hat{N}}} \nc{\hO}{{\hat{O}}}
\nc{\hP}{{\hat{P}}} \nc{\hR}{{\hat{R}}} \nc{\hS}{{\hat{S}}}
\nc{\hT}{{\hat{T}}} \nc{\hU}{{\hat{U}}} \nc{\hV}{{\hat{V}}}
\nc{\hW}{{\hat{W}}} \nc{\hX}{{\hat{X}}} \nc{\hZ}{{\hat{Z}}}

\nc{\hn}{{\hat{n}}}

\def\dg{\dagger}

\newcommand{\bra}[1]{\langle#1|}
\newcommand{\ket}[1]{|#1\rangle}

\def\Dbar{\leavevmode\lower.6ex\hbox to 0pt
{\hskip-.23ex\accent"16\hss}D}
\pdfoutput=1

\begin{document}

\large

\title{Quantum cost of dense coding and teleportation}

\date{\today}
\author{Xinyu Qiu}\email[]{xinyuqiu@buaa.edu.cn}
\affiliation{LMIB(Beihang University), Ministry of education, and School of Mathematical Sciences, Beihang University, Beijing 100191, China}
\author{Lin Chen}\email[]{linchen@buaa.edu.cn (corresponding author)}
\affiliation{LMIB(Beihang University), Ministry of education, and School of Mathematical Sciences, Beihang University, Beijing 100191, China}
\affiliation{International Research Institute for Multidisciplinary Science, Beihang University, Beijing 100191, China}

\begin{abstract}
The quantum cost is a key ingredient to evaluate the quality of quantum protocols from a practical viewpoint. We show that the quantum cost of $d$-dimensional
dense coding protocol is equal to $d+3$ when transmitting the classical message $(0,0)$, and that is equal to $d+4$ when transmitting other classical message. It appears linear growth with the dimension and thus makes sense for implementation. In contrast, the quantum cost of high-dimensional teleportation
protocols is equal to 13, which is the maximum value of the cost for the two-dimensional case. As an application, we establish the relation between the quantum cost and fidelity of dense coding protocols in terms of four typical noise scenario.
\end{abstract}
\maketitle

\section{Introduction}

  In the last decades, quantum communication  has been  a  prominent application of  quantum mechanics. Dense coding was firstly proposed by Bennett et al. in 1992 \cite{bennett1992communication}. It is a fascinating method to transmit two bits of classical information  using quantum resource like entanglement. One year later, quantum teleportation \cite{bennett1993teleport} was proposed to realize reliable transmission of  an unknown state. Recently, some applications of quantum communication such as secure quantum key distribution \cite{pirandola2017fundamental,paraiso2021photonic} have already been successfully deployed.  Teleportation of photonic qubits over long distances of up to 1,400 kilometers through an uplink channel has been reported \cite{ren2017ground}. A demonstration of teleportation from photons to the vibrations of nanomechanical resonators has been proposed \cite{harris2021qt, fiaschi2021optomechanical}. Superdense teleportation has been implemented by photon pairs to  communicate a specific class of single-photon ququarts with average fidelity of 87.0\% \cite{graham2015superdense}. The probabilistic implementation of a nonlocal operation using a nonmaximally entangled state is developed \cite{chen2005probabilistic}.
  Dense coding and teleportation are generalized  with quantum states in high-dimensional Hilbert space \cite{zhang2020study,fonseca2019highdim}, as the qudit states with higher  robustness to noise  improve the channel capacity and the information security. The relation
between quantum error-correcting codes (QECCs) in heterogeneous systems and quantum information masking is indicated \cite{shi2021k}.  A scheme for teleportation of arbitrarily high-dimensional photonic quantum states are proposed, and the averaged fidelity is calculated to be 75\% in the current experiment \cite{luo2019qt}. Since the high-dimensional unitary operations are more difficult to implement in physics experiments, it is necessary to measure the implementation  cost of quantum protocols  by calculating the quantum cost.  Quantum cost of an arbitrary gate was first introduced by Barenco et al. in 1995 \cite{barenco1995elementary}. Generally,  the quantum cost of a circuit is the sum of the cost of  each gate used in designing the circuit.  The more the quantum cost is, the more complexity the execution of the circuit has. Quantum cost is a common figure of merit to evaluate and compare different circuits and it is key to evaluate the quality of  protocols  both
theoretically and experimentally.  Since the quantum cost was first proposed,  efforts have been made to calculate the cost of unitary gates and quantum circuits.  A procedure has been presented to optimize distributed quantum circuits in terms of teleportation cost for a predetermined partitioning \cite{zomo2018optimizing}.
 An efficient method has been proposed to reduce the number of teleportation requirement based on the commuting of quantum gates \cite{daei2021improving}. In a recent work, the quantum cost of teleporting a single qubit message among six different entangled channels has been calculated and compared \cite{2108.02641}. However, the situation may become more complex when we consider the quantum cost of higher dimensional teleportation protocol. Besides, the quantum cost of high-dimensional dense coding protocol has not been studied yet.

  In this paper, we analyze the quantum cost of $d$-dimensional dense coding and  teleportation protocol. We obtain that the quantum cost of $d$-dimensional
dense coding protocol is equal to $d+3$ when transmitting the classical message $(0,0)$, and that is equal to $d+4$ when transmitting other classical message.  As for the teleportation protocol, we generalize the two-dimensional Pauli-X gate to $d$-dimensional gates, which are implemented on the control qudit to recover the information. By adding appropriate such gates on the circuits, we obtain that the quantum cost of all high-dimensional teleportation protocols is equal to 13, which is the maximum value of the quantum cost for the two-dimensional case. The quantum protocol will finally need hardware to realize.  The implementation cost of a quantum circuit increases with its quantum cost. Our results show that the physical implementation cost of high-dimensional dense coding protocol presents a linear growth with the dimension. As an application of  the quantum cost, we show that the fidelities of dense coding decrease with the increase of its quantum cost. Since the quantum cost of $d$-dimensional teleportation remains 13 for any $d\geq 3$, we show that the demand for practical device to implement high-dimensional  teleportation remains  the same, without regard to the implementation cost for each gate influenced by the dimension.

Dense coding and teleportation protocols have been extended via multipartite entangled states, such as Greenberger-Horne-Zeilinger (GHZ) states \cite{xiong2016multiple, jiang2019cyclic}, W states \cite{ agrawal2006perfect}, cluster states \cite{liu2014qtofcluster} and genuine multiparticle entangled(GME) states \cite{yeo2006teleportation}. An explicit scheme has been designed  for the teleportation of an n-qubit quantum state. Its experimental realization is performed using 5-qubit superconductivity-based IBM quantum computer with high fidelity \cite{sisodia2017design}. A scheme of $1\rightarrow 2$ optimal universal asymmetric quantum telecloning for pure multiqubit
states is proposed \cite{chen2007asymmetric}. Since the multipartite entangled states can be regarded as bipartite states, our study of bipartite high-dimensional dense coding and teleportation will have influence on the multipartite case.

  The rest of this paper is organized as follows. In Sec. \ref{sec:priliminary},  we introduce some basic concepts and list the basic gates used in this paper. Based on that, we decompose the non-basic gates into basic gates and calculate the quantum cost of each gate. In Sec. \ref{sec:densecoding} and \ref{sec:qt},  we  calculate the quantum cost of high-dimensional dense coding and  teleportation  protocol, respectively. We show the application of quantum cost in Sec. \ref{sec:application}. Finally, we conclude in Sec. \ref{sec:conclusions}.

\section{Preliminaries}
\label{sec:priliminary}
In this section, we review some basic concepts, decompose the non-basic gates and calculate their quantum cost. In Sec. \ref{sec:quantumcost}, we introduce the concept of quantum cost and its computation. In Sec. \ref{sec:basicgate}, we show  the basic gates used in this paper. In Sec.\ref{sec:preparegate}, we  show the decomposition of non-basic gates. Based on that, we obtain the quantum cost of each gate used in the dense coding and teleportation protocols.

\subsection{Quantum cost}
\label{sec:quantumcost}
 The quantum cost of a circuit is obtained by adding up the cost of each gate in the circuit. An arbitrary gate can be decomposed into several basic gates and the cost of  basic gates is considered to be a unit cost, regardless of their internal structure. That is to say,  we consider that the cost of a basic gate is one. If a gate can be decomposed into $n$ basic gates, then the quantum cost of the gate is equal to $n$.  When we refer to the  quantum cost  of a  protocol, we mean the quantum cost of the corresponding circuit. Mohammadi  and Eshghi \cite{Mohammadi2009reversible} have proposed two prescriptions for the calculation of quantum cost:
 \begin{enumerate}
 \item Implement a circuit using only the quantum primitive gates and count them;
 \item Synthesise a circuit using the gates whose quantum cost is specified. Add up the quantum cost of each gate in the circuit and obtain the  total quantum cost of the circuit.
 \end{enumerate}
 In this paper, we shall follow  prescription 2. We consider a gate  primitive if it maps decomposable states to decomposable states, which means that the primitive gate can not generate entanglement. Obviously, some gates used in the dense coding and teleportation protocol should have the ability to generate entanglement and hence they are not primitive, for example, the  CNOT gate.  Besides, some gates used in the two protocols can be prepared by the gates whose quantum cost is specified.
\subsection{Basic gate}
\label{sec:basicgate}
  Barenco et al. have considered all $2\times 2$ single-qubit gate and the CNOT gate as the basic gates in the two-dimensional case  and shown that we can realize the control-operations by at most six basic gates \cite{barenco1995elementary}. They have shown that the CNOT gate along with single-qubit gates may be assembled to do any quantum computation. The basic qubit gate can be extended to the basic qudit gate. JL Brylinski and R Brylinski \cite{2001Universal} have proposed that the collection of all one-qudit gates together with a two-qudit imprimitive gate is universal, i.e. every $n$-qudit  gate can be approximated with arbitrary accuracy by this collection of gates. Hence, all the single-qudit gate and a two-qudit imprimitive gate are the basic gates in the $d$-dimensional case.

 In this paper, we consider the single-qudit unitary gates, i.e. $H_d$,  $H_d^\dg$, $U_{mn,d}$, $P_{k,d}$, $Z_{k,d}$ gate and the two-qudit imprimitive CNOT gate as the basic gates. That is to say, the two-qudit gates should be decomposed with the help of CNOT gate. The expression of the two-dimensional basic gates are shown in \ref{sec2dbasicgates}, and that of $d$-dimensional basic gates are given in Sec. \ref{secddbasicgate}.

\subsubsection{Two-dimensional basic gate}
\label{sec2dbasicgates}
We list some two-dimensional basic gates used in this paper:
\begin{enumerate}
\item The two-dimensional Hadamard gate
\begin{eqnarray}
H_2=\frac{1}{\sqrt{2}}\bma1&1\\1&-1\ema;
\end{eqnarray}
\item The CNOT gate
\begin{eqnarray}
 U_{CN,2}=&&\ket{0,0}\bra{0,0}+\ket{0,1}\bra{0,1}+\ket{1,1}\bra{1,0}+\ket{1,0}\bra{1,1};
\end{eqnarray}
\item The  Pauli-X and Pauli-Z  matrix
\begin{eqnarray}
\s_X=\bma 0&1\\1&0\ema,  \s_Y=\bma 0&-i\\i&0\ema,\s_Z=\bma 1&0\\0&-1\ema.
\end{eqnarray}
\end{enumerate}
The quantum cost of these two-dimensional basic gates is equal to one.

\subsubsection{$d$-dimensional basic gate, $d>2$}
\label{secddbasicgate}
We show the $d$-dimensional basic gates which can be considered as the generalization of two-dimensional basic gates.

 We set $\og=e^{\frac{2\pi i}{d}}$. The $d$-dimensional Hadamard gate is
\begin{eqnarray}
 H_d=\frac{1}{\sqrt{d}}\sum_{x,y=0}^{d-1}\og^{xy}\ket{x}\bra{y}.
\end{eqnarray}
When $d=2$, we have $H_d=H_2$. The quantum cost of the $H_d$ gate is equal to one.

The $H_d^\dg$ gate is used to recover the message in dense coding protocol. The expression of this gate is 
\begin{eqnarray}
	H_d^\dg=\frac{1}{\sqrt{d}}\sum_{x,y=0}^{d-1}\og^{x(d-y)}\ket{x}\bra{y}.
\end{eqnarray}
It is a single-qudit basic gate, and its quantum cost is equal to one. 

The CNOT gate performs the transformation $\ket{a,b}\rightarrow\ket{a,a\oplus b}$, where "$\oplus$" denotes sum modulo $d$.  The expression of this gate is,
\begin{eqnarray}
 U_{CN,d}=\sum_{x,y=0}^{d-1}\ket{x,y\oplus x}\bra{x,y}.
\end{eqnarray}
When $d=2$, we have $ U_{CN,d}= U_{CN,2}$. The quantum cost of the CNOT gate is equal to one.

We show its function on the controlled qudit, which will be used later. If the control qudit is $\ket{k}$, then the CNOT gate performs the operation $X_{k,d}$ on the controlled qudit,
\begin{eqnarray}
X_{k,d}\ket{j}=\ket{j\oplus k},
\end{eqnarray}
where
\begin{eqnarray}
\label{defgatexkd}
X_{k,d}=\sum_{s=0}^{d-1}\ket{s\oplus k}\bra{s}.
\end{eqnarray}

 The  $U_{mn,d}$ gates are used to implement the operation corresponding to the classical message to be transmitted in dense coding protocol. The expression of this kind of gate is
\begin{eqnarray}
	\label{defUmn}
	U_{mn,d}=\sum_{u=0}^{d-1}\og^{mu}\ket{u}\bra{n\oplus u},
\end{eqnarray}
where $m,n=0,1,...,d-1$. 

 The following $P_{k,d}$ gates play an important role in dense coding and quantum teleportation protocol,
\begin{eqnarray}
    P_{k,d} =\sum_{s=0}^{d-1}\ket{s}\bra{k\oplus (d-s)}.
\end{eqnarray}
The quantum cost of $P_{k,d}$ gates is equal to one. Note that when $d=2$, we have $P_{0,2}=I,P_{1,2}=\s_X$. So it is a generalization of two-dimensional basic gate.
\subsection{Some gates prepared by the basic gates}
\label{sec:preparegate}
We  prepare the Control-Z and  CNOT$^\dg$ gate by the basic gates and show their quantum cost. They are used in the dense coding and teleportation protocol later.

The Control-Z gate is a two-qudit gate. It can be prepared with the help of a CNOT gate and two Hadamard gates, i.e.
\begin{eqnarray}
U_{CZ,d}=(I\otimes H_d)U_{CN,d}(I\otimes H_d).
\end{eqnarray}
The gate can be decomposed to three basic gates. The  quantum cost of it is equal to three.

If the control qudit is $\ket{k}$,  then the Control-Z gate performs the   operation $Z_{k,d}$ of the controlled qudit, where
\begin{eqnarray}
Z_{k,d}=&&H_dX_{k,d}H_d, (k=0,1,...,d-1).
\end{eqnarray}
One can verify that
\begin{eqnarray}
\label{defgatezkd}
Z_{k,d}=\sum_{j=0}^{d-1}\og^{kj}\ket{j}\bra{d-j}.\nonumber
\end{eqnarray}
 When $d=2$, we have $Z_{0,2}=I_2,Z_{1,2}=\s_Z$. The two-qudit Control-Z gate is the generalization of the two-qubit Control-Z gate.

 The  CNOT$^\dg$ gate is used to recover the message in dense coding. It is used to implement the operation $\ket{a,b}\rightarrow \ket{a,b\oplus (d-1)a)}$. For all the two-qudit gates, the only basic gate is CNOT gate. Hence, the CNOT$^\dg$ gate should be decomposed with the help of CNOT gate. One can obtain that it can be prepared by $d-1$ CNOT gates, i.e.
\begin{eqnarray}
U_{CN,d}^\dg=\sum_{x,y=0}^{d-1}\ket{x,y\oplus(d-1) x}\bra{x,y}=(U_{CN,d})^{d-1}.
\end{eqnarray}
The quantum cost of the CNOT$^\dg$ gate is equal to $d-1$.
\section{Quantum cost of  dense coding}
\label{sec:densecoding}
\begin{figure}[h]
	\center{\includegraphics[width=8cm]  {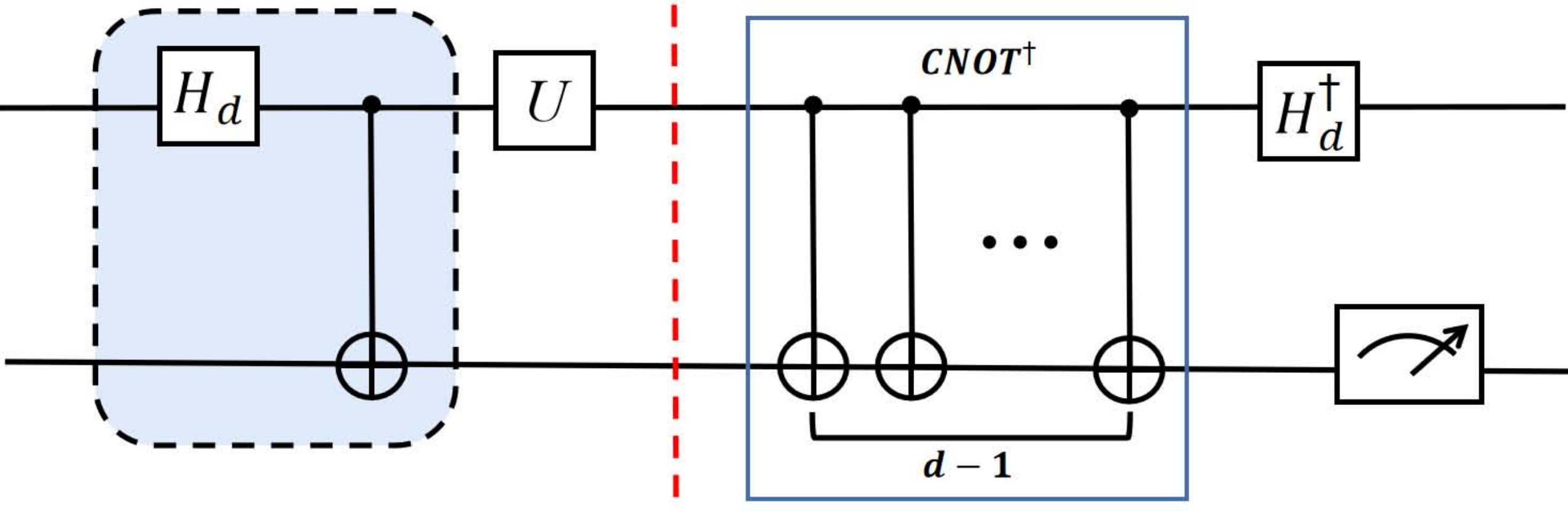}}
	\caption{The  protocol for the $d$-dimensional dense coding. The left two gates are used to prepare Bell states $\ket{\phi_1}=\frac{1}{\sqrt{d}}\sum_{k=0}^{d-1}\ket{k,k}$ with the qudit $\ket{0,0}$. The gates after the dashed line are used by Bob to recover the message. }
\label{figure:dcd1}
\end{figure}

In this section, we show the quantum cost of high-dimensional dense coding protocol. Suppose Alice and Bob share the $d$-dimensional Bell channel $\ket{\phi_1}=\frac{1}{\sqrt{d}}\sum_{k=0}^{d-1}\ket{k,k}$. Alice has the first qudit and Bob has the second qudit. Now, Alice wants to send two dits of classical message to Bob. The protocol is shown in FIG. \ref{figure:dcd1}.

The two-dit classical message may be one of the elements in the set $\{(0,0),(0,1),...,(d-1,d-1)\}$. If Alice wants to send the classical message $(m,n)$ to Bob, the gate $U_{mn,d}$ implements appropriate operations on her qudit.  The operation of the gate on $\ket{\phi_1}$ is
\begin{eqnarray}
(U_{mn,d}\otimes I){\ket{\phi_1}}=(\sum_{u=0}^{d-1}\og^{mu}\ket{u}\bra{n\oplus u})\otimes I)\ket{\phi_1}=\ket{\phi_{md+n+1}},
\end{eqnarray}
where $m,n=0,1,...,d-1$. They are shown in TABLE \ref{tabel:operationU} in detail.

Then Alice transmits her qubit to Bob. Bob tries to recover the message by two gates $H_d^\dg$ and CNOT$^\dg$.
One can verify that
\begin{eqnarray}
(H_d^\dg\otimes I)U_{CN,d}^{\dg}\ket{\phi_{md+n+1}}=\ket{m,n}.
\end{eqnarray}
Finally, Bob performs the measurement to his qudits and obtain the  classical message. The quantum cost of the final measurement is equal to one.

Now we show the quantum cost of the protocol in FIG.\ref{figure:dcd1}.  The key is to analyze the quantum cost of the gate $U_{mn,d}$. When $d=2$, we have $H_2^\dg=H_2$, CNOT$_2^\dg$=CNOT$_2$.  Alice sends one of the  classical message  $(0,0),(0,1),(1,0),(1,1)$ to Bob.
When Alice sends the classical message $(m,n)$ to Bob, the quantum cost of the dense coding protocol($D_{mn,2}$) is
\begin{eqnarray}
D_{mn,2}=&&D(H_2)+D(U_{mn,2})+D(CN_2)+D(M)\\\nonumber
=&&2\times1+DC(U_{mn,2})+2\times1+1\\
=&&D(U_{mn,2})+5,\nonumber
\end{eqnarray}
where $D(X)$ is the total quantum cost of all the $X$ gate used in this dense coding protocol, and $D(M)=1$ is the cost of the final measurement.
Based on the corresponding gate $U_{mn,2}$ shown  in TABLE \ref{table:costdc2}, we have
 \begin{eqnarray}
 D(U_{00,2})=0, D(U_{01,2})=1, D(U_{10,2})=1, D(U_{11,2})=1.
 \end{eqnarray}
 We obtain the quantum cost of the two-dimensional protocol(shown in TABLE \ref{table:costdc2}).

\begin{table}[h]
\caption{The classical message and corresponding operation implemented by the $U_{mn,2}$ gate on the channel $\ket{\ph_1}$.   The last two columns contain the quantum cost and the number of sorts of gates used in the quantum circuit for the two-dimensional case.}
\label{table:costdc2}
\begin{tabular}{|c|c|c|c|c|}	
\hline
 Classical Message&Operation $U_{mn,2}$&$(U_{mn,2}\otimes I)\ket{\ph_1}$&Quantum Cost&Sorts of Basic Gates\\
\hline
$(0,0)$&$U_{00,2}=I$&$\ket{\ph_1}=\frac{1}{\sqrt{2}}(\ket{0,0}+\ket{1,1})$&5&2\\
\hline
$(0,1)$&$U_{01,2}=\s_X$&$\ket{\ph_2}=\frac{1}{\sqrt{2}}(\ket{0,1}+\ket{1,0})$&6&3\\
\hline
$(1,0)$&$U_{10,2}=\s_Z$&$\ket{\ph_3}=\frac{1}{\sqrt{2}}(\ket{0,0}-\ket{1,1})$&6&3\\
\hline
$(1,1)$&$U_{11,2}=i\s_Y$&$\ket{\ph_4}=\frac{1}{\sqrt{2}}(\ket{0,1}-\ket{1,0})$&6&3\\
\hline
\end{tabular}
\end{table}

 We consider the case $d>2$ for any $d$. In FIG. \ref{figure:dcd1}, we see that $H_d$, CNOT, $U_{mn,d}$, CNOT$^\dg$, $H_d^\dg$ gate are used once in the protocol.  In Sec. \ref{secddbasicgate}, we have shown that $D(CN_d^\dg)=d-1$. When Alice wants to send classical message $(m,n)$ to Bob, the quantum cost of the $d$-dimensional protocol($ D_{mn,d}$) is
 \begin{eqnarray}
 D_{mn,d}=&&D(H_d)+D(CN_d)+D(U_{mn,d})+D(CN_d^\dg)+D(H_d^\dg)+D(M)\nonumber\\
=&&1+1+D(U_{mn,d})+(d-1)+1+1\\
=&&D(U_{mn,d})+d+3,\nonumber
 \end{eqnarray}
where $m,n=0,1,...,d-1$. Note that  $U_{00,d}=I$ and $U_{mn,d}\neq I$, for $m\neq 0$ or $n\neq 0$. We have
\begin{eqnarray}
D(U_{mn,d})=\begin{cases}
0&\mbox{if $m,n=0$,}\\
1&\mbox{otherwise.}
\end{cases}
\end{eqnarray}
Hence, for any $d\geq 2$, we have
\begin{eqnarray}
\label{defdccost}
D_{mn,d}=\begin{cases}
d+3&\mbox{if $m,n=0$,}\\
d+4&\mbox{otherwise.}
\end{cases}
\end{eqnarray}
The $H_d$, CNOT, $U_{mn,d}$, $H_d^\dg$ gates are the basic gates used in this protocol. The unbasic  CNOT$^\dg$ gate is prepared by $d-1$ CNOT gates. Hence,  four kinds of basic gates are used in the  circuit for dense coding protocol.

\begin{table}[h]
\caption{The classical message and corresponding operation implemented by the $U_{mn,d}$ gate on the channel $\ket{\phi_1}$. The last column contains the quantum cost of the $d$-dimensional dense coding protocol, $d\geq 3$. }
\label{tabel:operationU}
\begin{tabular}{|c|c|c|c|}	
\hline Classical Message&$(U_{mn,d}\otimes I)\ket{\phi_1}$& Quantum Cost\\
\hline $(0,0)$&$\ket{\phi_1}=\frac{1}{\sqrt{d}}(\ket{0,0}+\ket{1,1}+...+\ket{d-1,d-1})$&$d+3$\\
\hline $(0,1)$&$\ket{\phi_2}=\frac{1}{\sqrt{d}}(\ket{0,1}+\ket{1,2}+...+\ket{d-1,0})$&$d+4$\\
\hline \vdots&\vdots&\vdots\\
\hline $(0,d-1)$&$\ket{\phi_d}=\frac{1}{\sqrt{d}}(\ket{0,d-1}+\ket{1,0}+...+\ket{d-1,d-2})$&$d+4$\\
\hline \vdots&\vdots&\vdots\\
\hline $(d-1,0)$&$\ket{\phi_{d^2-d+1}}=\frac{1}{\sqrt{d}}(\ket{0,0}+\og^{d-1}\ket{1,1}+...+\og\ket{d-1,d-1})$&$d+4$\\
\hline $(d-1,1)$&$\ket{\phi_{d^2-d+2}}=\frac{1}{\sqrt{d}}(\ket{0,1}+\og^{d-1}\ket{1,2}+...+\og\ket{d-1,0})$&$d+4$\\
\hline \vdots&\vdots&\vdots\\
\hline $(d-1,d-1)$&$\ket{\phi_{d^2}}=\frac{1}{\sqrt{d}}(\ket{0,d-1}+\og^{d-1}\ket{1,0}+...+\og\ket{d-1,d-2})$&$d+4$\\
\hline
\end{tabular}
\end{table}
~\\
\section{Quantum cost of teleportation}
\label{sec:qt}
In this section, we show the  quantum cost of $d$-dimensional teleportation protocol. The single-qudit quantum message is written as $\ket{M_d}=\sum_{j=0}^{d-1}\a_j\ket{j}$, where $\a_j\in\bbC$ and $\sum_{j=0}^{d-1}|\a_j|^2=1$. Alice and Bob share the $d$-dimensional maximally entangled state  $\ket{\phi_1}=\frac{1}{\sqrt{d}}\sum_{k=0}^{d-1}\ket{k,k}$ as the channel. So they are
in the state
\begin{eqnarray}
\ket{\xi_d}=\ket{M_d}\otimes\ket{\phi_1}=(\sum_{j=0}^{d-1}\a_j\ket{j})_1\otimes \frac{1}{\sqrt{d}}\sum_{k=0}^{d-1}\ket{k,k})_{2,3}.
\end{eqnarray}
The qudit 1 and 2 of the combined state belong to Alice and the qudit 3 belongs to Bob. Alice and Bob implement the operations on their qudits.  The final state is shown as follows,
\begin{flalign}
\label{deffsd}
&(H_d\otimes I\otimes I)(U_{CN,d}\otimes I)\ket{\xi_d}=\sum_{m,n=0}^{d-1}\frac{\ket{m,n}_{1,2}}{d}\left[\sum_{j=0}^{d-1}\a_j\og^{mj}\ket{n\oplus(d-j)}_3\right].
\end{flalign}

Based on the state given in (\ref{deffsd}), we obtain that when Alice's measurement is $\ket{m,n}$, Bob should apply the local unitary operation $U_{mn}^{(0,0)}=Z_{d-m}X_{d-n}$ to his qudit, where
  \begin{eqnarray}
  U_{m,n}^{(0,0)}=\sum_{s=0}^{d-1}\og^{ms}\ket{d-s}\bra{s\oplus n}.
  \end{eqnarray}
  Finally, Bob performs the measurement  and completes the teleportation process. The quantum cost of the final measurement is equal to one.

\begin{figure}[htp]
	\center{\includegraphics[width=8cm]  {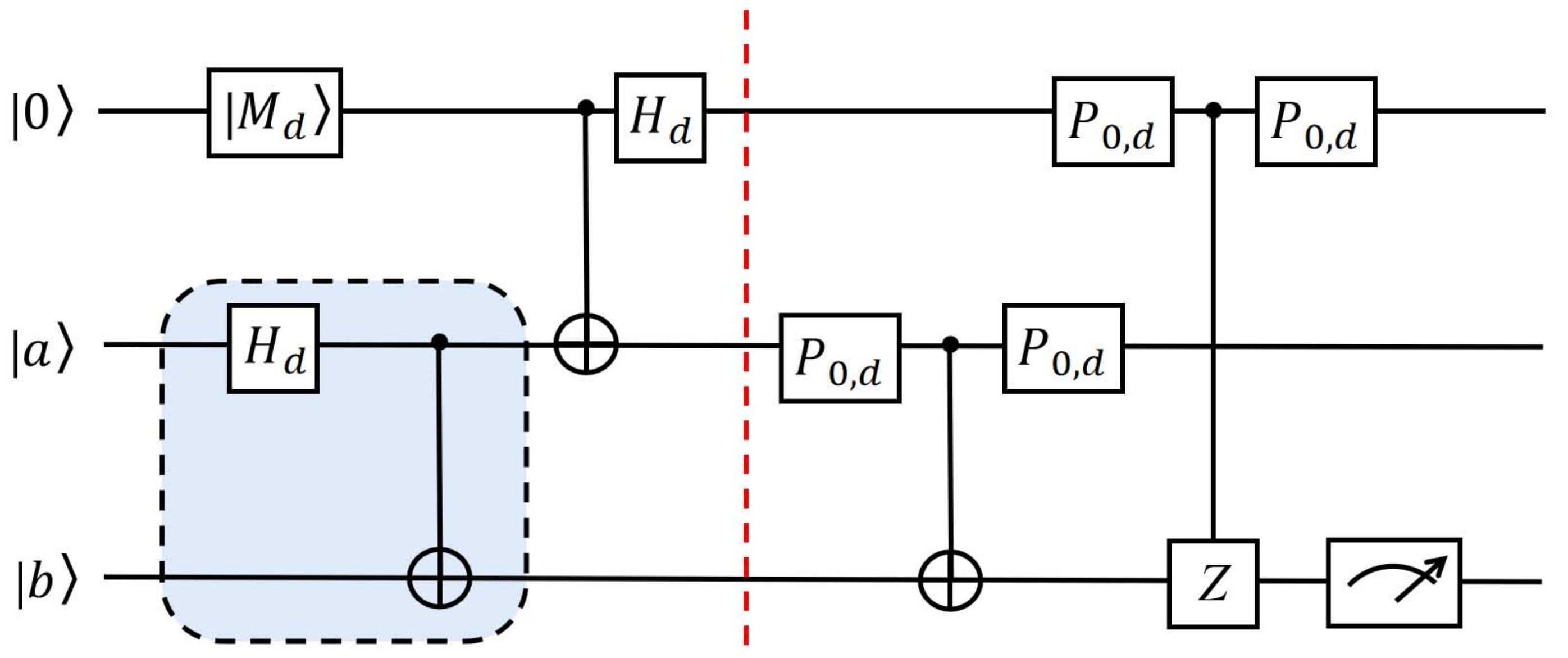}}
	\caption{The teleportation protocol via the $d$-dimensional Bell state $\ket{\phi_1}=\frac{1}{\sqrt{d}}\sum_{k=1}^{d-1}\ket{k,k}$. The unitary gates to recover the message are shown after the  dashed line.  Three kinds of basic gates are used in the circuit: $H_d$, CNOT, and $P_{0,d}$ gate.}
\label{figure:qcd1}
\end{figure}

 We analyze the quantum cost of the teleportation protocol via $d$-dimensional channel $\ket{\phi_1}$ for $d>2$. In FIG.\ref{figure:qcd1}, we show that two $H_d$ gates, three CNOT gates, four $P_{0,d}$ gates and one Control-Z gate are used in this protocol. In Sec. \ref{sec:basicgate}, we have shown that the quantum cost of the $H_d$, CNOT, $P_{0,d}$ gate is equal to one, and the cost of the Control-Z gate is equal to three.    We add up the quantum cost of each gate and the final measurement and obtain that the quantum cost of the $d$-dimensional protocol($T_d$) is
\begin{eqnarray}
T_d=&&T(H_d)+T(CN_d)+T(P_{0,d})+T(CZ_d)+T(M)\\
=&&2\times 1+3\times 1+4\times 1+1\times 3+1=13,\nonumber
\end{eqnarray}
where $T(X)$ is the total quantum cost of the $X$ gate used in this protocol, and $T(M)=1$ is the cost of the final measurement.
  Three sorts of basic gates($H_d$, $P_{0,d}$, CNOT gate)  are used in the teleportation protocol.

\begin{figure}[htp]
	\center{\includegraphics[width=8cm]  {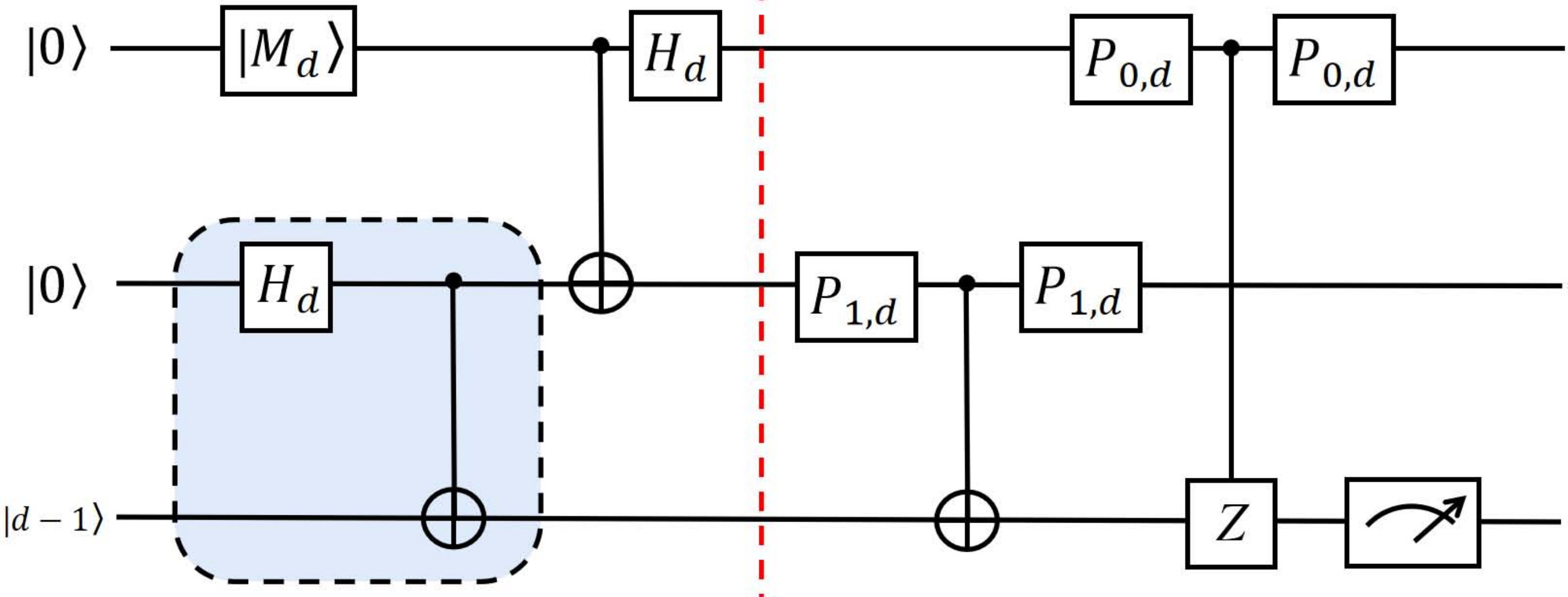}}
	\caption{The teleportation protocol via the $d$-dimensional Bell state $\ket{\phi_d}$. The unitary gates to recover the message are shown after the  dashed line.  Four kinds of basic gates are used in the circuit: $H_d$, CNOT, $P_{0,d}$ and $P_{1,d}$ gate.}
\label{figure:qcdd}
\end{figure}

Next we analyze the quantum cost of the teleportation protocol via other $d$-dimensional Bell channels. In the quantum circuit shown in FIG.\ref{figure:qcd1}, we have $\ket{a},\ket{b}\in\{\ket{0},\ket{1},...,\ket{d-1}\}$. Choosing all combinations of  $\ket{a}$ and $\ket{b}$, we obtain $d^2$ kinds of  Bell channels,
\begin{eqnarray}
\ket{\phi_{ad+b+1}}=\frac{1}{\sqrt{d}}\sum_{x=0}^{d-1}\og^{xa}\ket{x,b\oplus x}.
\end{eqnarray}
Via the channel $\ket{\phi_{ad+b+1}}$, when Alice's measurement result is $\ket{m,n}$, the corresponding operation that Bob should apply is
\begin{eqnarray}
U_{m,n}^{(a,b)}=Z_{a\oplus (d-m)}X_{(d-b)\oplus(d-n)},
\end{eqnarray}
they are shown in TABLE \ref{tabel:operationd} in detail. Hence, the gates implement on the first and second qudit are $P_{a,d}$ and $P_{d-b,d}$. They transform the control qudit of Control-Z and CNOT gate into appropriate value, respectively. The type of $P_{k,d}$ gates is the only difference between the teleportation protocols via different channels. The number of $P_{k,d}$ gates in quantum circuits via different Bell channels are all equal to four. For example, comparing FIG. \ref{figure:qcd1} and FIG. \ref{figure:qcdd}, we see that the difference between the teleportation protocols via $\ket{\phi_1}$ and $\ket{\phi_d}$ is the type of $P_{k,d}$ gates implement on the control qudit of CNOT gate. The quantum cost of the teleportation protocol via all the Bell channels $\ket{\phi_u},(u=1,2,...,d^2)$ is equal to  13. One or two kinds of $P_{k,d}$ gates, $H_d$ gates and CNOT gates are used in the circuit. Hence, three or four kinds of basic gates are used in the teleportation protocol via all the Bell channels.

When $d=2$, the  quantum cost is different from the case $d>2$, as we have $P_{0,2}=I_2$, which is used four times in FIG.\ref{figure:qcd1}.  Hence, the quantum cost of the teleportation protocol via the two-dimensional channel $\ket{\varphi_1}=\frac{1}{\sqrt{2}}(\ket{00}+\ket{11})$ is
$$T_2=T_d-T(P_{0,d})=13-4=9.$$
That is the reason why the quantum cost of teleportation protocol via the two-dimensional channels varies from 9 to 13. The quantum cost of teleportation protocols via four two-dimensional channels and Bob's recover operations are given in TABLE \ref{tabel:chooseab2}. It is in agreement  with previous analysis for the case of dimension two in \cite{2108.02641}.


\begin{table}[htp]
\caption{Controlled operations that Bob should apply to recover the message $\ket{M_d}$ via the $d$-dimensional Bell channels. When the control qubit is $\ket{k}$, the CNOT and Control-Z gate performs the operations $X_{k,d}$ and $Z_{k,d}$ on Bob's qudit. Here $X_k$ and $Z_k$ represent $X_{k,d}$ and $Z_{k,d}$  given in (\ref{defgatexkd}) and (\ref{defgatezkd}), ($k=0,1,...,d-1$).}
\label{tabel:operationd}
\begin{tabular}{|c|c|c|c|c|c|c|c|c|c|}	
\hline Alice's measurement&$\ket{\phi_1}$&$\ket{\phi_2}$&\ldots&$\ket{\phi_d}$&\ldots
&$\ket{\phi_{d^2-d+1}}$&$\ket{\phi_{d^2-d+2}}$&\ldots&$\ket{\phi_{d^2}}$\\
\hline
$\ket{0,0}$&$Z_0X_0$ & $Z_0X_{d-1}$&\ldots&$Z_0X_1$ &\ldots &$Z_{d-1}X_0$ &$Z_{d-1}X_{d-1}$ &\ldots&$Z_{d-1}X_1$\\
\hline
$\ket{0,1}$&$Z_0X_{d-1}$ & $Z_0X_{d-2}$&\ldots&$Z_0X_0$ &\ldots&$Z_{d-1}X_{d-1}$ &$Z_{d-1}X_{d-2}$ &\ldots&$Z_{d-1}X_0$\\
\hline
\vdots&\vdots& \vdots& &\vdots & &\vdots &\vdots & &\vdots\\
\hline
$\ket{0,d-1}$&$Z_0X_1$ & $Z_0X_0$&\ldots &$Z_0X_2$ &\ldots &$Z_{d-1}X_1$ &$Z_{d-1}X_0$ &\ldots&$Z_{d-1}X_2$\\
\hline
\vdots&\vdots& \vdots& &\vdots & &\vdots &\vdots & &\vdots\\
\hline
$\ket{d-1,0}$&$Z_1X_0$ & $Z_1X_{d-1}$&\ldots&$Z_1X_1$ &\ldots &$Z_0X_0$ &$Z_0X_{d-1}$ &\ldots&$Z_0X_1$\\
\hline
$\ket{d-1,1}$&$Z_1X_{d-1}$ & $Z_1X_{d-2}$&\ldots &$Z_1X_0$ &\ldots &$Z_0X_{d-1}$ &$Z_0X_{d-2}$ &\ldots&$Z_0X_0$\\
\hline
\vdots&\vdots& \vdots& &\vdots & &\vdots &\vdots & &\vdots\\
\hline
$\ket{d-1,d-1}$&$Z_1X_1$ & $Z_1X_0$&\ldots &$Z_1X_2$ &\ldots &$Z_0X_1$ &$Z_0X_0$ &\ldots&$Z_0X_2$\\
\hline
\end{tabular}
\end{table}

 \begin{table}[htp]
\caption{We set $d=2$ in TABLE \ref{tabel:operationd} to  obtain the operations that Bob should apply for the two-dimensional case. The last column contains the quantum cost of the two-dimensional teleportation protocol via different channels. Here $X_k$ and $Z_k$ represent $X_{k,2}$ and $Z_{k,2}$, $k=0,1$. }
\label{tabel:chooseab2}
\begin{tabular}{|c|c|c|c|c|c|c|c|}	
\hline$\ket{a}$&$\ket{b}$& Two-dimensional Bell States&\multicolumn{4}{|c|}{Alice's Measurement}&Quantum Cost\\
\cline{4-7}
 & & &$\ket{0,0}$&$\ket{0,1}$&$\ket{1,0}$&$\ket{1,1}$&\\
\hline $\ket{0}$&$\ket{0}$&$\ket{\varphi_1}=\frac{1}{\sqrt{2}}(\ket{0,0}+\ket{1,1}$&$Z_0X_0=I_2$&$Z_0X_1=\s_X$&$Z_1X_0=\s_Z$&$Z_1X_1=\s_Z\s_X$&9\\
\hline $\ket{0}$&$\ket{1}$&$\ket{\varphi_2}=\frac{1}{\sqrt{2}}(\ket{0,1}+\ket{1,0}$&$Z_0X_1=\s_X$&$Z_0X_0=I_2$&$Z_1X_1=\s_Z\s_X$&$Z_1X_0=\s_Z$&11\\
\hline $\ket{1}$&$\ket{0}$&$\ket{\varphi_3}=\frac{1}{\sqrt{2}}(\ket{0,0}-\ket{1,1}$&$Z_1X_0=\s_Z$&$Z_1X_1=\s_Z\s_X$&$Z_0X_0=I_2$&$Z_0X_1=\s_X$&11\\
\hline $\ket{1}$&$\ket{1}$&$\ket{\varphi_4}=\frac{1}{\sqrt{2}}(\ket{0,1}-\ket{1,0}$&$Z_1X_1=\s_Z\s_X$&$Z_1X_0=\s_Z$&$Z_0X_1=\s_X$&$Z_0X_0=I_2$&13\\
\hline
\end{tabular}
\end{table}
\section{Applications}
\label{sec:application}
In this section, we introduce an application of the quantum cost. As we all know, some unavoidable interaction of the communication channel with the environment leads to the loss of accuracy of the protocol. In order to assess the reliability of a protocol, the fidelity  of the protocol is proposed \cite{Fonseca2019high, Schaetz2004quantum}. It gives the closeness between the ideal state Alice wants to send and the final state under noisy channel. By considering the channel under some classes of noise, we find the relation between the quantum cost and its fidelity. In  Sec. \ref{sec:qt} we have obtained  that the quantum cost of teleportation remains 13, regardless of the dimension. The fidelity of teleportation protocol has nothing to do with its quantum cost, as the fidelity is related to the dimension and error probability.  On the other hand, we will show that the fidelity of the dense coding protocol under four classes of noise  decreases with the increase of quantum cost. Hence, the quantum cost of a protocol is one of the indicators of its fidelity. The more gates or complicated gates we employ in a protocol, the more quantum cost will be, hence the less fidelity will be. It inspires us to optimize the protocol by decreasing the quantum cost of it as much as possible.

In order to calculate the fidelity of the dense coding protocol, we briefly introduce four classes of noise for $d$-dimensional case \cite{Fonseca2019high}. Suppose $d^2$ Weyl operators $U_{mn}$ are defined as
\begin{eqnarray}
U_{mn}=\sum_{j=0}^{d-1}\og^{jm}_d\ket{j\oplus n}\bra{j}.
\end{eqnarray}
In analogy to  the two-dimensional niose, four classes of noise and its
corresponding Kraus operators are shown as follows, where $p$ is the probability that the error occurs.
\begin{itemize}
  \item Dit-flip niose:  $E_{00}=\sqrt{1-p}U_{00}$, $E_{01}=\sqrt{\frac{p}{d-1}}U_{01}$,..., $E_{0,d-1}=\sqrt{\frac{p}{d-1}}U_{0,d-1}$.
  \item $d$-phase-flip noise: $E_{00}=\sqrt{1-p}U_{00}$, $E_{10}=\sqrt{\frac{p}{d-1}}U_{10}$,..., $E_{d-1,0}=\sqrt{\frac{p}{d-1}}U_{d-1,0}$.
  \item Dit-phase-flip noise: $E_{00}=\sqrt{1-p}U_{00}$, $E_{mn}=\frac{\sqrt{p}}{d-1}U_{mn}$, with $1\leq m,n\leq d-1$.
  \item Depolarizing noise: $E_{00}=\sqrt{1-\frac{d^2-1}{d^2}p}U_{00}$, $E_{mn}=\frac{\sqrt{p}}{d}U_{mn}$, with $0\leq m,n\leq d-1$, for $(m,n)\neq (0,0)$.
\end{itemize}
These four classes of noise contain the information about the effects of the system-environment interaction. Given an arbitrary system initially prepared in a state $\rho=\sum_{\vec{k}\vec{l}}\rho_{\vec{k}\vec{l}}\ket{\vec{k}}\bra{\vec{l}}$, for the number of subsystem $N$, $\vec{k}=(k_1,...k_N)$, $0\leq k_j\leq d-1$, the action of a set of Kraus operators $E_{\vec{k}\vec{l}}=E_{k_1l_1}\otimes...\otimes E_{k_Nl_N}$ transforms $\rho$ into $\rho'$. The evolution can be modeled by the trace preserving map $\rho\rightarrow\rho'=\sum_{\vec{k}\vec{l}}E_{\vec{k}\vec{l}}\rho E_{\vec{k}\vec{l}}^\dg$, where the $E_{\vec{k}\vec{l}}$'s  satisfy the completeness relation $\sum_{\vec{k}\vec{l}}E_{\vec{k}\vec{l}} E_{\vec{k}\vec{l}}^\dg=I$.

We take the dit-flip noise as an example to show the calculation of fidelity. Suppose Alice wants to send classical message $(m,n)$ to Bob. They share the channel $\rho=\ket{\phi_1}\bra{\phi_1}$, where $\ket{\phi_1}=\frac{1}{\sqrt{d}}\sum_{k=0}^{d-1}\ket{k,k}$. The  $U_{mn,d}$ gate shown in (\ref{defUmn}) implements the operation that Alice performs to her qudit.
In the noise-free environment, i.e. the channel $\rho$ is not affected by any kind of noise, the final state of Bob's two qudits is
\begin{eqnarray}
\rho_{mn}=(H_d^\dg\otimes I)U_{CN}^\dg(U_{mn}\otimes I)\rho(U_{mn}^\dg\otimes I)U_{CN}(H_d\otimes I).
\end{eqnarray}
The action of dit-flip  noise transforms the channel $\rho$ into $\rho'$:
\begin{eqnarray}
\rho'=&&\sum_{j,q=0}^{d-1}(E_{0j}\otimes E_{0q})\rho(E_{0j}\otimes E_{0q})^\dg
=\sum_{j,q=0}^{d-1}[(E_{0j}\otimes E_{0q})\ket{\phi_1}][\bra{\phi_1}(E_{0j}\otimes E_{0q})^\dg].
\end{eqnarray}
Hence, Bob's two qudits are transformed into the new state $\rho'_{mn}$, where
\begin{eqnarray}
\rho'_{mn}=&&(H_d^\dg\otimes I)U_{CN}^\dg(U_{mn}\otimes I)\rho'(U_{mn}^\dg\otimes I)U_{CN}(H_d\otimes I)\\
=&&\sum_{j,q=0}^{d-1} (H_d^\dg\otimes I)U_{CN}^\dg(U_{mn}\otimes I)[(E_{0j}\otimes E_{0q})\ket{\phi_1}][\bra{\phi_1}(E_{0j}\otimes E_{0q})^\dg](U_{mn}^\dg\otimes I)U_{CN}(H_d\otimes I).\nonumber
\end{eqnarray}
After some calculations, we obtain the operations in dense coding protocol
\begin{eqnarray}
(H_d^\dg\otimes I)U_{CN}^\dg(U_{mn}\otimes I)=\frac{1}{\sqrt{d}}\sum_{s,u,x=0}^{d-1}\og^{(m-x)s}\ket{x,u\oplus (d-s)}\bra{n\oplus s,u}.
\end{eqnarray}

For any $j,q=1,2,...,d-1$, we have
\begin{eqnarray}
\ket{\xi_{00}}=&&(H_d^\dg\otimes I)U_{CN}^\dg(U_{mn}\otimes I)(E_{00}\otimes E_{00})\ket{\phi_1}=\frac{1-p}{d}\sum_{s,x=0}^{d-1}\og^{(m-x)s}\ket{x,n},\\
\ket{\xi_{0q}}=&&(H_d^\dg\otimes I)U_{CN}^\dg(U_{mn}\otimes I)(E_{00}\otimes E_{0q})\ket{\phi_1}=\frac{\sqrt{p(1-p)}}{d\sqrt{d-1}}\sum_{s,x=0}^{d-1}\og^{(m-x)s}\ket{x,n\oplus q},\\
\ket{\xi_{j0}}=&&(H_d^\dg\otimes I)U_{CN}^\dg(U_{mn}\otimes I)(E_{0j}\otimes E_{00})\ket{\phi_1}=\frac{\sqrt{p(1-p)}}{d\sqrt{d-1}}\sum_{s,x=0}^{d-1}\og^{(m-x)s}\ket{x,n\oplus (d-j)},\\
\ket{\xi_{jq}}=&&
(H_d^\dg\otimes I)U_{CN}^\dg(U_{mn}\otimes I)(E_{0j}\otimes E_{0q})\ket{\phi_1}=\frac{p}{d(d-1)}\sum_{s,x=0}^{d-1}\og^{(m-x)s}\ket{x,n\oplus q\oplus (d-j)}.
\end{eqnarray}
Hence,
\begin{eqnarray}
\rho'_{mn}=\sum_{j,q=0}^{d-1}\ket{\xi_{j,q}}\bra{\xi_{j,q}}.
\end{eqnarray}
By considering an arbitrary classical message $(m,n)$ that Alice wants to send, the fidelity of dense coding under dit-flip noise($\cF_{F}$) is
\begin{eqnarray}
\cF_{F}=&&tr\{\ket{m,n}\bra{m,n}\rho'_{mn}\}=(1-p)^2+\frac{p^2}{d-1}.
\end{eqnarray}
In Sec. \ref{sec:densecoding}, we have obtained that when Alice wants to send $(m,n)$ to Bob, the quantum cost of the $d$-dimensional dense coding in (\ref{defdccost}). For simplicity, we consider that Alice sends the classic message other than $(0,0)$. The quantum cost is $D=D_{mn,d}=d+4$.

We establish the relation between the fidelity of dense coding protocol under dit-flip noise and its quantum cost,
\begin{eqnarray}
\cF_{F}=(1-p)^2+\frac{p^2}{D-5}.
\end{eqnarray}

\begin{figure}[htbp]
\centering
\subfigure[$\cF_F=\cF_P$]{
\includegraphics[width=5.5cm]{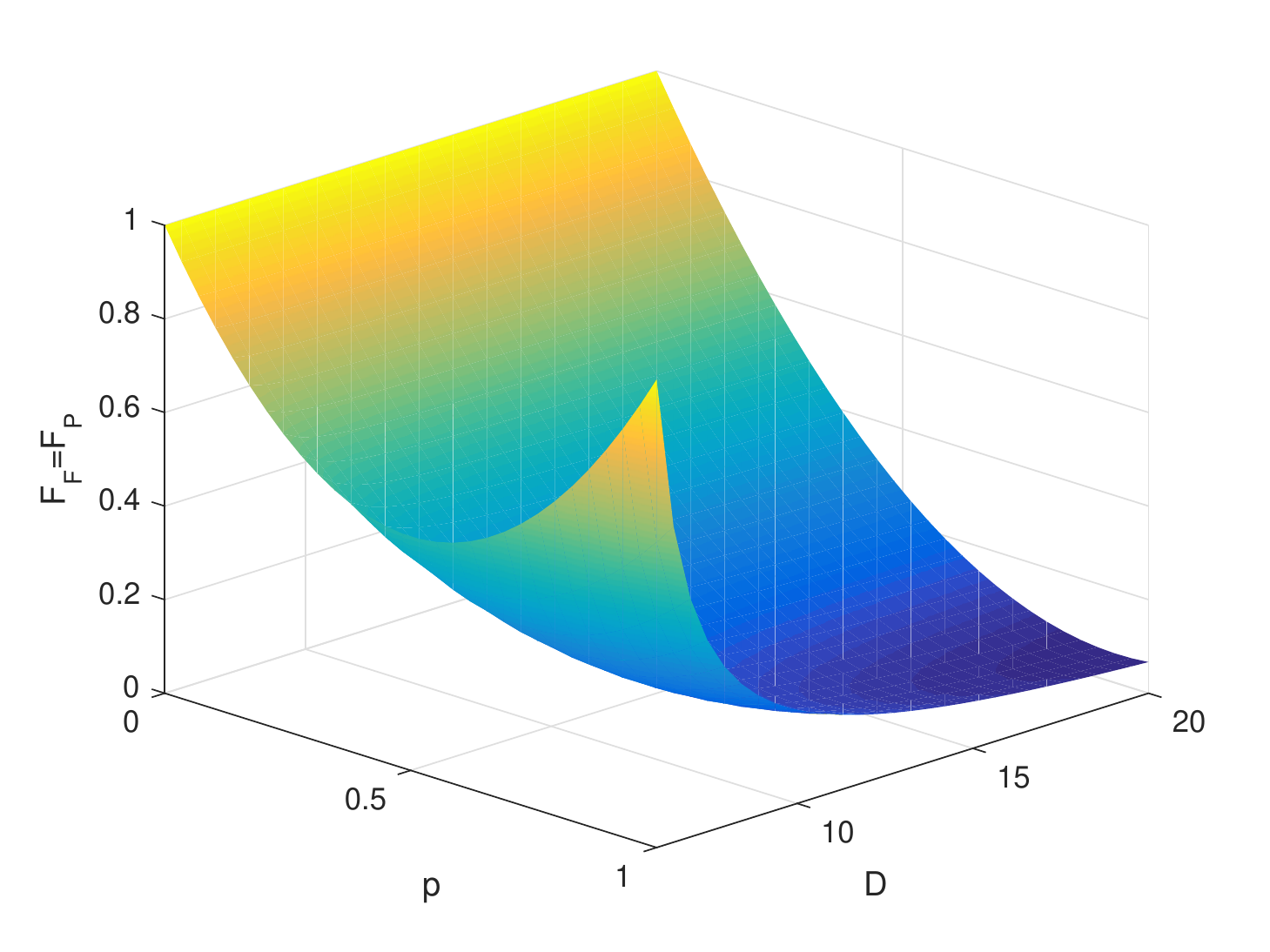}
}
\subfigure[$\cF_{FP}$]{
\includegraphics[width=5.5cm]{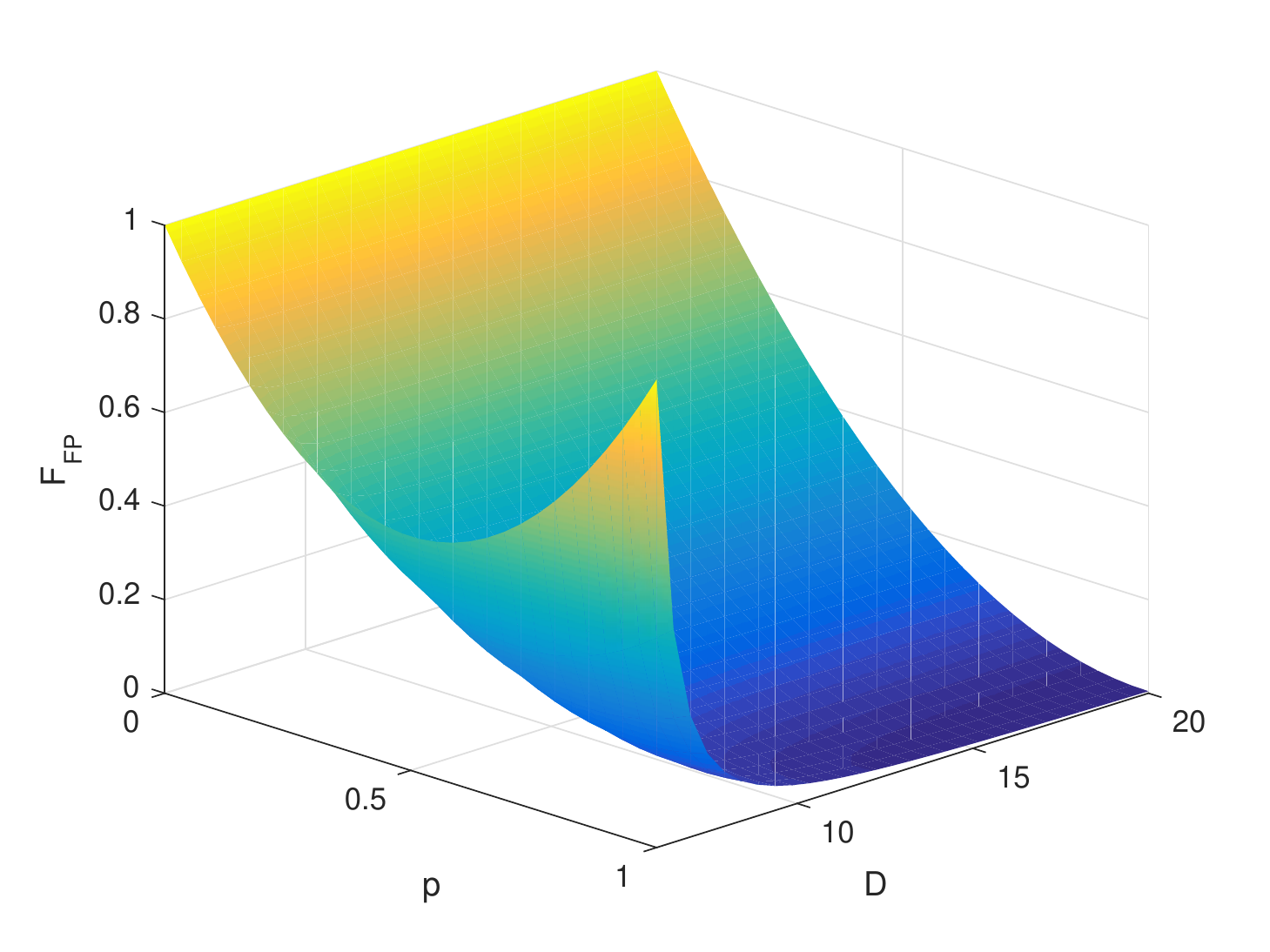}
}
\subfigure[$\cF_D$]{
\includegraphics[width=5.5cm]{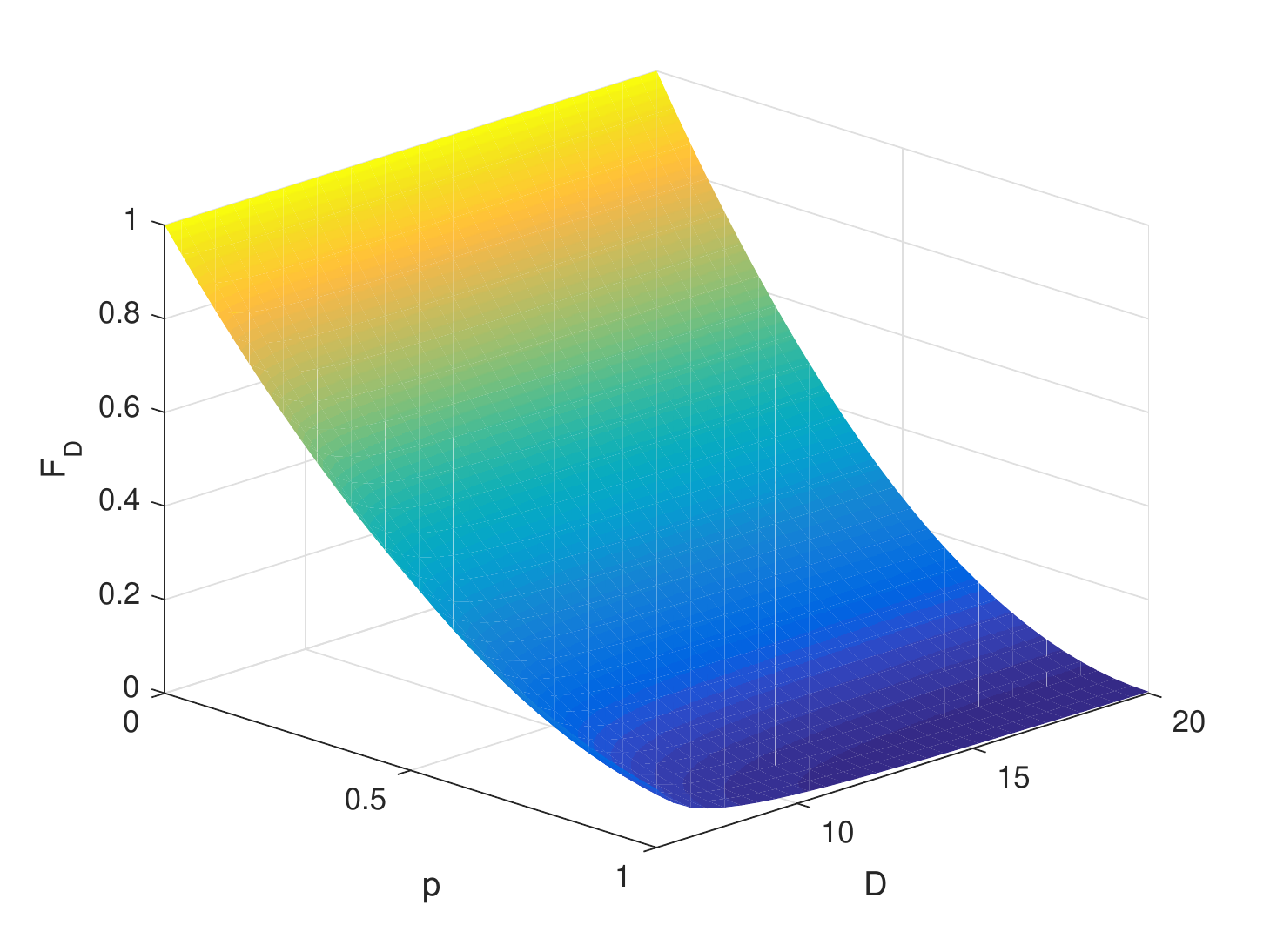}
}
\caption{ Fidelities of dense coding under a scenario in which the channel is affected by dit-flip($\cF_F$),$d$-phase flip($\cF_P$), dit-phase flip($\cF_{FP}$), and depolarizing noise($\cF_D$), respectively. The axes represent the fidelity, error probability $p$ and the quantum cost $D$, for $0\leq p\leq 1$ and $6\leq D \leq 20$.}
\label{figure:fidelity}
\end{figure}

By similar calculations, we obtain the following fidelities corresponding to $d$-phase flip noise($\cF_P$), dit-phase flip noise($\cF_{FP}$), and depolarizing noise($\cF_D$). We also establish the relation between the fidelities  under the three kinds of  noise and its quantum cost,
\begin{eqnarray}
\cF_P=&&(1-p)^2+\frac{p^2}{d-1}=(1-p)^2+\frac{p^2}{D-5}, \\
\cF_{FP}=&&(1-p)^2+\frac{p^2}{(d-1)^2}=(1-p)^2+\frac{p^2}{(D-5)^2},\\
\cF_{D}=&&(1-\frac{d^2-1}{d^2}p)^2+\frac{(d-1)^2p^2}{d^4}=\left[1-\frac{(D-4)^2-1}{(D-4)^2}p\right]^2+\frac{(D-5)^2p^2}{(D-4)^4}.
\end{eqnarray}
The fidelities of dense coding for noise scenario are plotted in FIG. \ref{figure:fidelity}. The figures show that the increase of quantum cost of dense coding protocol will result in the loss of its fidelity for the noise scenario. With the increase of $D$, the reduction rate of fidelity with respect to  $p$ becomes larger. Hence, decreasing the quantum cost  would be one of the useful strategies to improve the fidelity of the high-dimensional dense coding protocol.

\section{Conclusions}
\label{sec:conclusions}
We have analyzed the quantum cost of high-dimensional dense coding and teleportation protocol. Our results of the teleportation protocol have generalized the results recently shown for the two-dimensional case \cite{2108.02641}.  We have obtained that the quantum cost of $d$-dimensional dense coding protocol is equal to $d+3$ when transmitting the classical message $(0,0)$, and that is equal to $d+4$ when transmitting other classical message, showing a linear increase with the dimension. Four kinds of basic gates are used in the dense coding protocol. The quantum cost of high-dimensional teleportation remains 13, which is the maximum value of the quantum cost of two-dimensional case. Three or four kinds of basic gates are used in the teleportation protocol. As an application of our main result, we have been able to establish relations between the fidelity of dense coding protocol and its quantum cost. The more the quantum cost is, the less fidelity of the protocol will be for the four kinds of  noise scenario.

Many problems arising from this paper can be further explored. The quantum cost of other high-dimensional protocols may be obtained, and the relation between its fidelity and quantum cost can be established, for example, the two-step quantum direct communication protocol \cite{Deng2003two} and  the protocol for quantum secure direct communication with  superdense coding \cite{wang2005quantum}. It offers a strategy to improve the fidelity of protocols. Besides, we have studied the quantum cost of protocols by bipartite entangled states, which can be extended to  multipartite states. It is also left as an open problem whether there are relations between the quantum cost and entanglement cost in protocols.

\section*{ACKNOWLEDGMENTS}
We thank Yi Shen for a careful reading of the whole paper. The authors were supported by the NNSF of China (Grant No. 11871089) and the Fundamental Research Funds for the Central Universities (Grant No. ZG216S2005).
\bibliographystyle{unsrt}
\bibliography{quantum_cost_xinyu}
\end{document}